\documentclass{article}
\usepackage{waspaa17}
\usepackage{amsmath,amsfonts,graphicx,url,times}
\usepackage{color}

\usepackage{multirow}

\usepackage{algorithm}
\makeatletter
\newcommand{\removelatexerror}{\let\@latex@error\@gobble}
\makeatother
\usepackage[noend]{algpseudocode}
\usepackage{eqparbox}

\errorcontextlines\maxdimen
\makeatletter
    \newcommand*{\algrule}[1][\algorithmicindent]{\makebox[#1][l]{\hspace*{.5em}\thealgruleextra\vrule height \thealgruleheight depth \thealgruledepth}}%
\newcommand*{\thealgruleextra}{}
\newcommand*{\thealgruleheight}{.75\baselineskip}
\newcommand*{\thealgruledepth}{.25\baselineskip}

\newcount\ALG@printindent@tempcnta
\def\ALG@printindent{%
    \ifnum \theALG@nested>0
        \ifx\ALG@text\ALG@x@notext
        \else
            \unskip
            \addvspace{-1pt}
            \ALG@printindent@tempcnta=1
            \loop
                \algrule[\csname ALG@ind@\the\ALG@printindent@tempcnta\endcsname]%
                \advance \ALG@printindent@tempcnta 1
            \ifnum \ALG@printindent@tempcnta<\numexpr\theALG@nested+1\relax
            \repeat
        \fi
    \fi
    }%
\usepackage{etoolbox}
\patchcmd{\ALG@doentity}{\noindent\hskip\ALG@tlm}{\ALG@printindent}{}{\errmessage{failed to patch}}
\makeatother

\newbox\statebox
\newcommand{\myState}[1]{%
    \setbox\statebox=\vbox{#1}%
    \edef\thealgruleheight{\dimexpr \the\ht\statebox+1pt\relax}%
    \edef\thealgruledepth{\dimexpr \the\dp\statebox+1pt\relax}%
    \ifdim\thealgruleheight<.75\baselineskip
        \def\thealgruleheight{\dimexpr .75\baselineskip+1pt\relax}%
    \fi
    \ifdim\thealgruledepth<.25\baselineskip
        \def\thealgruledepth{\dimexpr .25\baselineskip+1pt\relax}%
    \fi
    \State #1%
    \def\thealgruleheight{\dimexpr .75\baselineskip+1pt\relax}%
    \def\thealgruledepth{\dimexpr .25\baselineskip+1pt\relax}%
}
\newif\ifjournal
\journalfalse 

\newif\ifarxiv
\arxivtrue 

\ifjournal
\title{Deep Recurrent Nonnegative Matrix Factorization Networks for Speech Separation
by Unfolding Iterative Thresholding
}
\else
\title{DEEP RECURRENT NMF FOR SPEECH SEPARATION \\
BY UNFOLDING ITERATIVE THRESHOLDING
}
\fi

\name{Scott Wisdom,$^{1}$
     Thomas Powers,$^{1}$
     James Pitton,$^{1,2}$ 
     Les Atlas,$^{1}$
     \thanks{
     Funded under ONR contract N00014-12-G-0078,
     delivery order
     0013.}
     }
\address{$^1$ Department of Electrical Engineering, University of Washington, Seattle, WA, USA\\
        $^2$ Applied Physics Laboratory, University of Washington, Seattle, WA, USA\\
        \{swisdom, tcpowers, pitton, atlas\}@uw.edu\\
}

\begin{document}

\ninept
\maketitle

\begin{sloppy}

\begin{abstract}
  In this paper, we propose a novel recurrent neural network architecture for speech separation. This architecture is constructed by unfolding the iterations of a sequential iterative soft-thresholding algorithm (ISTA) that solves the optimization problem for sparse nonnegative matrix factorization (NMF) of spectrograms. We name this network architecture \emph{deep recurrent NMF} (DR-NMF). The proposed DR-NMF network has three distinct advantages. First, DR-NMF provides better interpretability than other deep architectures, since the weights correspond to NMF model parameters, even after training. This interpretability also provides principled initializations that enable faster training and convergence to better solutions compared to conventional random initialization. Second, like many deep networks, DR-NMF is an order of magnitude faster at test time than NMF, since computation of the network output only requires evaluating a few layers at each time step. Third, when a limited amount of training data is available, DR-NMF exhibits stronger generalization and separation performance compared to sparse NMF and state-of-the-art long-short term memory (LSTM) networks. When a large amount of training data is available, DR-NMF achieves lower yet competitive separation performance compared to LSTM networks.
\end{abstract}


\section{Introduction}
\label{sec:intro}

Nonnegative matrix factorization (NMF) of spectrograms continues to be a popular and effective method for audio source separation, especially for separating speech from challenging real-world nonstationary background noise. Since NMF is based on a family of statistical models \cite{fevotte_nonnegative_2009}, it is easy to extend in many ways, such as adding sparsity \cite{eggert_sparse_2004, sun_universal_2013, le_roux_sparse_2015}, convolution \cite{ogrady_convolutive_2006}, dynamics across time, \cite{
fevotte_non-negative_2013, smaragdis_static_2014}, and phase awareness \cite{kameoka_complex_2009}. 
However, NMF has the disadvantage of needing to solve an optimization problem at test time, which can require computing many iterations of an optimization algorithm.

Deep neural networks (DNNs), especially recurrent neural networks (RNNs), have also been shown to be very effective for audio source separation, as long as a large supervised dataset is available to optimize the weights of the network \cite{huang2014deep, weninger_discriminatively_2014, weninger_speech_2015}. DNNs are powerful models that can learn complicated nonlinear mappings from large amounts of data, and as a result tend to outperform NMF models on the speech separation problem when provided with enough data. Additionally, DNNs are fast at test time, requiring the computation of only a few nonlinear layers. However, despite their excellent performance and unlike NMF, DNNs lack interpretability, which prevents diagnosis of training difficulties and leaves trial-and-error as the only means of constructing and improving DNNs. 

In this paper, we propose a new type of recurrent neural network architecture that combines the advantages of NMF and deep recurrent neural networks: the \emph{deep recurrent NMF} (DR-NMF) network. Like NMF, our proposed DR-NMF network is interpretable, in that its weights directly correspond to parameters of an underlying statistical model.
However, since the DR-NMF network can be trained with backpropagation, it is able to take advantage of larger amounts of data than NMF. Interestingly, the DR-NMF network essentially consists of a deep feedforward rectified linear unit (ReLU) network at each time step, where the current input data is connected to each layer. DR-NMF is also fast at test time, since it can it run in online mode and only a few layers need to be evaluated at each time step. In practice, we find that DR-NMF networks achieve competitive separation performance compared to state-of-the-art long-short term memory (LSTM) networks, outperforming LSTM networks when an order of magnitude less training data is used.

The paper is organized as follows. First, we describe related work. Then we review required background, including sparse NMF for speech separation and
ISTA.
We then explain the unfolding of ISTA for sparse NMF to a DR-NMF network and its training. Finally, we present and discuss results on the CHiME2 dataset.

\section{RELATION TO PRIOR WORK}
\label{sec:relation}

DR-NMF is an instance of the very general procedure of deep unfolding \cite{hershey_deep_2014}, which is a method of converting inference algorithms for statistical models into novel deep network architectures that can then be trained using backpropagation on a supervised dataset.

DR-NMF is not the first time NMF has been combined with deep networks. Le Roux et al. \cite{le_roux_deep_2015} proposed unfolding multiplicative updates for sparse NMF, which they called deep NMF. Deep NMF networks were shown to outperform deep feedforward networks for speech separation using fewer trainable parameters. We go beyond this work by exploiting recurrence between the NMF coefficients of adjacent frames. Also, instead of multiplicative updates, we unfold a different optimization algorithm for the sparse NMF problem, ISTA, which is an accelerated gradient descent method that is fast and easier to optimize with backpropagation.

Gregor and LeCun \cite{gregor_learning_2010} were the first to unfold the ISTA algorithm and learn its parameters, which they called learned ISTA (LISTA). Rolfe and LeCun \cite{rolfe_discriminative_2013} added a discriminative classification term to the LISTA cost function.
Borgerding et al.\ \cite{borgerding_amp-inspired_2017} unfolded the approximate message passing algorithm (AMP) for sparse coding and observed improved performance compared to learned ISTA and fast ISTA (FISTA).
Kamilov and Mansour \cite{kamilov_learning_2016, kamilov_learning_2016-1} learned optimal nonlinear thresholding functions for ISTA and FISTA. Kamilov et al.\ \cite{kamilov_learning_2017} learned improved proximal filters for unfolded FISTA.
This paper
goes beyond
these works by considering sequential data and exploiting time-recurrent 
connections between consecutive ISTA optimizations.
While we only consider unfolding a sequential version of ISTA in this paper, ideas from these prior works, such as unfolding FISTA and AMP or learning better thresholding functions, could be combined with our novel approach.

Recently, we proposed unfolding sequential ISTA (SISTA) into a novel and interpretable stacked RNN architecture called the SISTA-RNN \cite{wisdom_interpretable_2016, wisdom_building_2017}. The DR-NMF network in this paper is a modified SISTA-RNN applied to solve sparse NMF for speech separation. DR-NMF is different from the SISTA-RNN in two ways: a nonnegativity constraint is placed on the sparse coefficients, and the Gaussian penalty used by SISTA between reconstructed observations at consecutive time steps is removed.


\section{BACKGROUND}
\label{sec:statistical}

Assume that $D$ samples $x_d=y_d+v_d$, $d=1..D$, of a noisy audio signal are observed, where $y_{1:D}$ is a desired clean speech signal and $v_{1:D}$ is additive noise.
From these noisy samples, the $F\times T$ nonnegative magnitude or power spectrogram matrix ${\bf X}$ is computed from the complex-valued short-time Fourier transform (STFT) matrix ${\bf X}_\mathbb{C}$ of $x$: ${\bf X}_\mathbb{C}=\mathrm{STFT}\{x_{1:D}\}$ and ${\bf X}=|{\bf X}_\mathbb{C}|$.

\subsection{NMF for speech separation}

NMF assumes that ${\bf X}$ can be decomposed into the product of a $F\times N$ elementwise nonnegative dictionary ${\bf W}$ and a $N \times T$ elementwise nonnegative activation matrix ${\bf H}$: ${\bf X} \approx \hat{\bf X} = {\bf W}{\bf H}$, where $F$ is the number of frequency bins, $N$ is the number of NMF basis vectors, and $T$ is the number of spectrogram frames. Usually, the activation matrix ${\bf H}$ is constrained to be sparse, which promotes parsimonious representations and avoids trivial solutions \cite{eggert_sparse_2004, le_roux_sparse_2015}.

Training the dictionary ${\bf W}$ consists of solving the 
problem
\begin{equation}
    \begin{aligned}
    & \underset{{\bf W}\in\mathcal{W}, {\bf H} \geq 0}{\text{minimize}}
    & & D_\beta({\bf X} || {\bf W}{\bf H})
    + \lambda_1 \|{\bf H}\|_1\\
    \end{aligned}
    \label{eq:opt_nmf}
\end{equation}
where the set
$\mathcal{W}$
consists of all elementwise nonnegative matrices that have unit-norm columns.
The function
$D_\beta({\bf X} || \hat{\bf X})$
is the beta-divergence summed across time and frequency: $D_\beta({\bf X} || \hat{\bf X})=\sum_{f,t} d_\beta(X_{f,t} || \hat{X}_{f,t})$ \cite{fevotte_nonnegative_2009}, which
\ifjournal
\else
corresponds to the summed squared error for $\beta=2$. This paper will focus on the $\beta=2$ case, but the method in this paper can be used with other values of $\beta$.
\fi

To separate speech, a dictionary ${\bf W}^{(y)}$ is first trained on clean speech. Then a noise dictionary ${\bf W}^{(v)}$ is trained by using the concatenated overall dictionary ${\bf W}=[{\bf W}^{(y)}, {\bf W}^{(v)}]$, where the clean speech dictionary ${\bf W}^{(y)}$ remains fixed and only the noise dictionary ${\bf W}^{(v)}$ and the overall activation matrix ${\bf H}=[{\bf H}^{(y)}; {\bf H}^{(v)}]$ is updated (the notation ``$;$'' indicates row concatenation). To infer the separated speech at test time, the problem (\ref{eq:opt_nmf}) is solved with just ${\bf H}$ as a variable, keeping the overall dictionary ${\bf W}$ fixed.

To reconstruct the separated speech signal, a time-frequency 
filter, or mask, is computed with elements between $0$ and $1$:
\begin{equation}
    \hat{\bf M}
    =
    \frac{ \hat{\bf Y} }{ \hat{\bf Y} + \hat{\bf V} },
    \label{eq:mask}
\end{equation}
where division is elementwise, $\hat{\bf Y}={\bf W}^{(y)} {\bf H}^{(y)}$ is the estimated spectrogram of clean speech, and $\hat{\bf V}={\bf W}^{(v)}{\bf H}^{(v)}$ is the estimated spectrogram of noise. This mask is applied to the complex STFT matrix ${\bf X}_\mathbb{C}$ and the estimated speech signal is the inverse STFT:
$\hat{y}=\mathrm{STFT}^{-1}\{\hat{\bf M}\odot{\bf X}_\mathbb{C}\}$.

\subsection{Iterative soft-thresholding algorithm (ISTA) for NMF}

The conventional optimization algorithm used to solve problem (\ref{eq:opt_nmf}) is alternating multiplicative updates \cite{lee_algorithms_2001}. However, the convergence of multiplicative updates can be slow and backpropagating through these updates is challenging \cite{le_roux_deep_2015}, so we consider another optimization algorithm that is not often applied to NMF: the iterative soft-thresholding algorithm, commonly known as ISTA.

ISTA is an accelerated 
gradient descent algorithm to
solve the problem
\begin{equation}
    \begin{aligned}
    & \underset{{\bf h}}{\text{minimize}}
    & & f({\bf x}, {\bf h})
    + \lambda_1 g({\bf h}),\\
    \end{aligned}
    \label{eq:opt_ista}
\end{equation}
where $f$ is a smooth function and $g$ is a nonsmooth function \cite{chambolle_nonlinear_1998, daubechies_iterative_2004}. ISTA enjoys a $1/K$ rate of convergence, which improves over the $1/\sqrt{K}$ convergence of simple first-order gradient descent on (\ref{eq:opt_ista}) \cite{beck_fast_2009}, where $K$ is the number of iterations. The ISTA algorithm for $f({\bf x}, {\bf h})=\frac{1}{2}\|{\bf x}-{\bf W}{\bf h}\|_2^2$ and $g({\bf h})=\|{\bf h}\|_1$ is shown in algorithm \ref{alg:ista} (see \cite[Appendix B]{wisdom_interpretable_2016} for a derivation), where $1/\alpha$ is a step size and $\mathrm{soft}_b({\bf z})$ of a vector ${\bf z}$ denotes application of the following 
operation with real-valued threshold $b$ to each element $z_n$ of ${\bf z}$:
\begin{equation}
\label{eq:soft}
    \mathrm{soft}_{b}(z_n)
    =
    \frac{z_n}{|z_n|}\mathrm{max}(|z_n|-b,0).
\end{equation}
Note that when ${\bf h}$ has a nonnegativity constraint, as in problem (\ref{eq:opt_nmf}), the soft-thresholding operation is one-sided: $\mathrm{soft}_b(z_n|h_n\geq0)=\max(z_n-b, 0)$, which is often called a ReLU \cite{glorot_deep_2011}.

\begin{figure}[t]
\removelatexerror
\vspace{-7.5pt}
\begin{algorithm}[H]
 \caption{Basic iterative soft-thresholding algorithm (ISTA)}
 \label{alg:ista}
 \begin{algorithmic}[1]
 \renewcommand{\algorithmicrequire}{\textbf{Input:}}
 \Require
 observations ${\bf x}$, dictionary ${\bf W}$,
 initial coefficients ${\bf h}^{(0)}$
  \For{$k = 1$ to $K$}
    \State
    $
    \makebox[0pt][l]{{\bf z}}\phantom{{\bf h}^{(k)}}
    \leftarrow
    ({\bf I}-\frac{1}{\alpha}{\bf W}^T{\bf W}){\bf h}^{(k-1)}
    +
    \frac{1}{\alpha}{\bf W}^T
    {\bf x}
    $
    \State
    $
    {\bf h}^{(k)}
    \leftarrow
    \mathrm{soft}_{\lambda/\alpha}
    \left(
    {\bf z}
    \right)
    $
  \EndFor \\
 \Return ${\bf h}^{(K)}$
 \end{algorithmic}
 \end{algorithm}
 \vspace{-30pt}
 \end{figure}
 
 \begin{figure}[t]
\removelatexerror
\begin{algorithm}[H]
 \caption{Warm start ISTA}
 \label{alg:sista}
 \begin{algorithmic}[1]
 \renewcommand{\algorithmicrequire}{\textbf{Input:}}
 \Require
 observations ${\bf x}_{1:T}$, dictionary ${\bf W}$, initial coefficients ${\bf h}_0^{(K)}$
  \For{$t = 1$ to $T$}
    \State
    $
    {\bf h}_t^{(0)}
    \leftarrow
    {\bf h}_{t-1}^{(K)}
    $
    \Comment{~~\# {\it warm start from $t-1$}}
  \For{$k = 1$ to $K$}
    \State
    $
    \makebox[0pt][l]{{\bf z}}\phantom{{\bf h}_t^{(k)}}
    \leftarrow
    ({\bf I}-\frac{1}{\alpha}{\bf W}^T{\bf W}){\bf h}_t^{(k-1)}
    +
    \frac{1}{\alpha}{\bf W}^T
    {\bf x}_t
    $
    \State
    $
    {\bf h}_t^{(k)}
    \leftarrow
    \mathrm{soft}_{\lambda/\alpha}
    \left(
    {\bf z}
    \right)
    $
  \EndFor
  \EndFor \\
 \Return ${\bf h}^{(K)}$
 \end{algorithmic}
 \end{algorithm}
 \vspace{-25pt}
 \end{figure}
 
The ISTA algorithm can be used at test time to solve the NMF problem for ${\bf H}$ by running ISTA independently for each time frame. That is, in algorithm \ref{alg:ista}, ${\bf x}$ will be the $t$th column of ${\bf X}$, ${\bf W}$ will be the trained dictionary, and ${\bf h}$ will be the $t$th column of the solution ${\bf H}$. We will elect to use a fixed number of ISTA iterations, which we will refer to as $K$.

However, running ISTA independently on each time frame neglects potential correlation between adjacent frames. Thus, we make ISTA sequential, or recurrent in time, by allowing the ISTA iterations for frame $t$ to use the output of the previous frame $t-1$ as initialization, providing a ``warm start''. That is, ${\bf h}_t^{(0)}$ is set equal to ${\bf h}_{t-1}^{(K)}$, which is the output of $K$ ISTA iterations from frame $t-1$.
This procedure is described in algorithm \ref{alg:sista}. More sophisticated types of recurrence could also be incorporated, such as modeling the sequence of sparse activations ${\bf h}_{1:T}$ with a dynamical system \cite{fevotte_non-negative_2013}, which is a promising avenue for future work.

\section{UNFOLDING ISTA TO DEEP RECURRENT NMF}
\label{sec:unfolding}

In this section, we unfold warm start ISTA (algorithm \ref{alg:sista}) for sparse NMF that we described in the last section, which results in a DR-NMF network. The right panel of figure \ref{fig:comp_graph} illustrates the architecture of the DR-NMF network. Note that the warm starts, which are initialization of frame $t$ with the solution from frame $t-1$, manifest as recurrent connections across time.

\begin{figure}[t]
  \centering
  \centerline{
  \includegraphics[width=0.91\columnwidth]{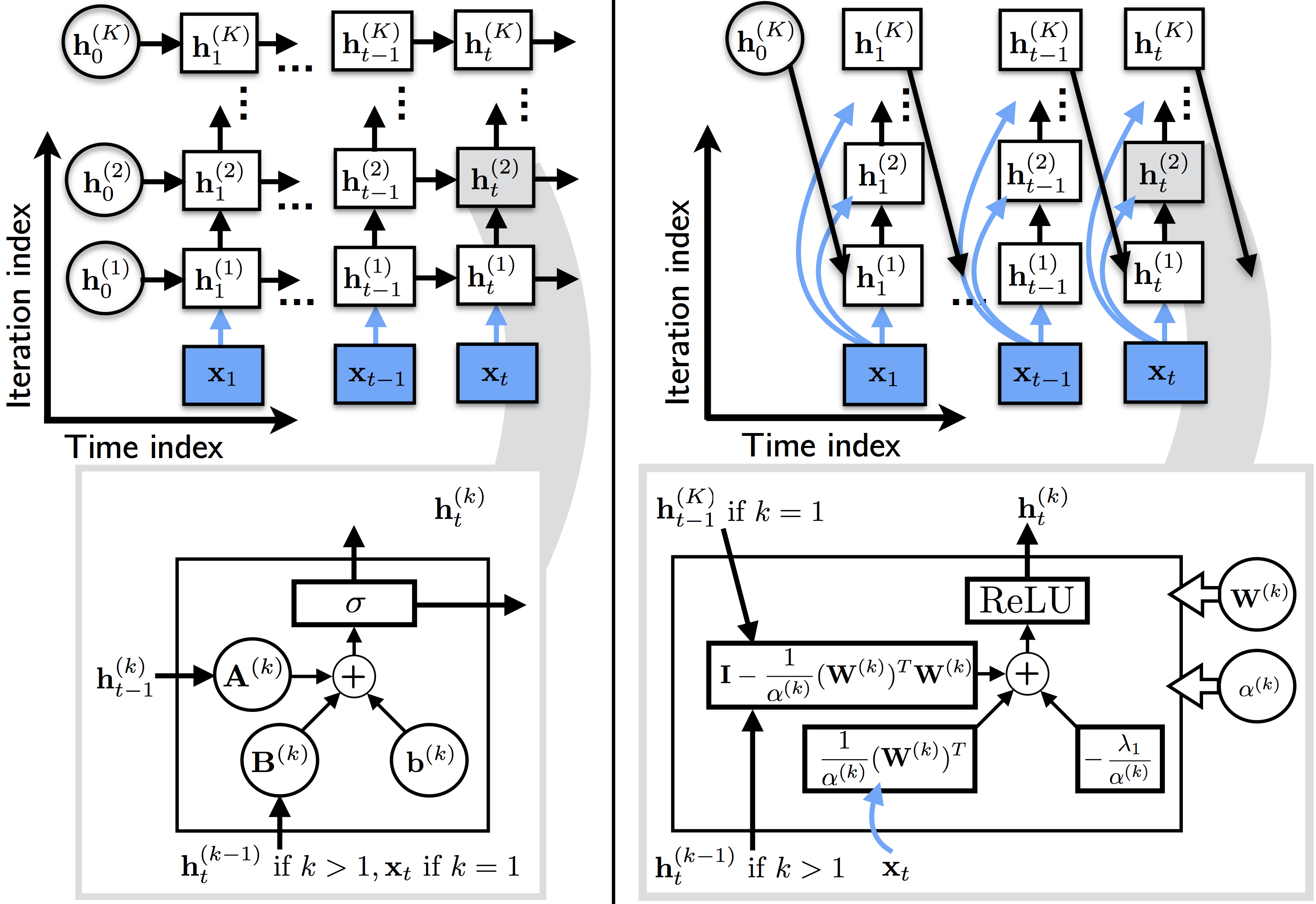}
  }
  \vspace{-10pt}
  \caption{Left panel: architecture of conventional stack of RNNs, corresponding to equation (\ref{eq:rnn}). Right panel: architecture of DR-NMF network, which is algorithm \ref{alg:sista} unfolded into a computational graph. Circles indicate trainable weights.
  }
  \label{fig:comp_graph}
  \vspace{-10pt}
\end{figure}

Since the nonlinear activation function for nonnegative ISTA is a ReLU, we can see that the unfolded warm start ISTA algorithm is essentially a conventional stack of RNNs that use a ReLU activation function, with two significant differences. First, for each time step $t$, the input is connected to each node in the deep stack. Second, the only recurrent connection between deep stacks at adjacent time frames is from the top node ${\bf h}_{t-1}^{(K)}$ at $t-1$ to the bottom node ${\bf h}_t^{(1)}$ at $t$.
For comparison, the left panel of figure \ref{fig:comp_graph} shows
a conventional stacked RNN, where the output of the $k$th RNN layer is
\begin{equation}
{\bf h}_t^{(k)}
= \sigma_{{\bf b}^{(k)}}(
    {\bf A}^{(k)}{\bf h}_{t-1}^{(k)}
    + {\bf B}^{(k)}{\bf h}_t^{(k-1)}),
\label{eq:rnn}
\end{equation}
where $\sigma_{\bf b}$ is an activation function with bias ${\bf b}$, e.g. the LSTM activation function \cite{hochreiter_long_1997}, and for the first layer ($k=1$), ${\bf h}_t^{(0)}={\bf x}_t$.

To train a stacked RNN or DR-NMF network with a supervised dataset $\{{\bf X}_i, {\bf Y}_i\}_{i=1:I}$ with $I$ examples, we solve the 
problem
\begin{equation}
    \begin{aligned}
    & \underset{\theta}{\text{minimize}}
    & & \frac{1}{I}\sum_{i=1}^I
    \ell \left(
        {\bf Y}_i, q_\theta({\bf X}_i)
        \right),\\
    \end{aligned}
    \label{eq:opt_train}
\end{equation}
where $\ell$ is a training loss function, $q_\theta$ is the neural network output, and $\theta$ are the weights of the network. We use stochastic gradient descent, where backpropagation through $q_\theta$ is used to compute the gradients with respect to the trainable parameters $\theta$.

For the speech separation problem, an input example ${\bf X}$ is the $F \times T$ magnitude spectrogram of a noisy audio file, while the output ${\bf Y}$ is the $F \times T$ magnitude spectrogram of the corresponding clean speech signal that we are trying to predict with the neural network. For our experiments, we use the signal approximation cost function \cite{weninger_speech_2015}. This cost function assumes that the network estimates a $F\times T$ masking matrix $\hat{\bf M}$ using (\ref{eq:mask}) and multiplies this mask elementwise with the noisy input, $q_\theta({\bf X}) = \hat{\bf M} \odot {\bf X}$. Then the signal approximation loss is the mean squared error between the true clean spectrogram ${\bf Y}$ and the estimated clean spectrogram $\hat{\bf M}\odot{\bf X}$:
\begin{equation}
    \ell({\bf Y}, q_\theta({\bf X}))
    =
    \sum_{f,t}
    (Y_{f,t} - \hat{M}_{f,t}X_{f,t})^2,
    \label{eq:loss}
\end{equation}
which corresponds to maximizing the signal-to-noise ratio (SNR) of magnitude spectra in the time-frequency domain.

Since DR-NMF corresponds to an optimization algorithm that solves the NMF optimization problem (\ref{eq:opt_nmf}), we can initialize DR-NMF using sparse NMF.
The overall training procedure is described by the following steps:
\begin{enumerate}
    \item Train clean speech dictionary ${\bf W}^{(y)}$ on clean speech audio using well-done sparse NMF multiplicative updates \cite{le_roux_sparse_2015}.
    \item Train noise dictionary ${\bf W}^{(v)}$ on noisy speech audio by building the overall dictionary ${\bf W}=[{\bf W}^{(y)}, {\bf W}^{(v)}]$ and only updating ${\bf W}^{(v)}$ using well-done sparse NMF multiplicative updates \cite{le_roux_sparse_2015}.
    \item Initialize DR-NMF network with the learned dictionary ${\bf W}$ and ISTA optimization parameters $\alpha$ and ${\bf h}_0$.
    \item Train the DR-NMF parameters $\theta=\{{\bf W}^{(1:K)}, \alpha^{(1:K)}, {\bf h}_0\}$, where the dictionary ${\bf W}$ and inverse step size $\alpha$ are untied across layers, by solving the problem (\ref{eq:opt_train}) using stochastic gradient descent with the signal approximation loss (\ref{eq:loss}).
\end{enumerate}

When initializing the DR-NMF network, the ISTA inverse step size $\alpha$ must be chosen appropriately to ensure that a fixed number $K$ of iterations achieves sufficient decrease in the sparse NMF objective function (\ref{eq:opt_nmf}).
\ifjournal
\else
Since $\alpha$ corresponds to the Lipschitz smoothness of $f$, if we assume the maximum inner product between two different columns of ${\bf W}$ is upper-bounded by $\delta$, then
    \ifarxiv
$\alpha \geq 2(1 + \delta (N-1))^2$.
    \else
$\alpha \geq 1 + \delta (N-1)$.
    \fi
\fi
We found that
\ifarxiv
$\alpha=50$ for $N=100$ and $\alpha=400$ for $N=1000$ work well.
\else
choosing $\alpha=N/4$ works well in
practice.
\fi

To ensure nonnegativity of the DR-NMF weights, we optimize the logs of the weights through an elementwise $\exp$ function. For example, for the nonnegative weight ${\alpha}$, we optimize $\widetilde{\alpha}$, which is initialized with $\log{(\epsilon+{\alpha})}$, and use $\exp{(\widetilde{{\alpha}})}$ for the model weight. To maintain nonnegativity and unit-norm columns of ${\bf W}$, we optimize $\widetilde{\bf W}$, which is initialized with $\log{(\epsilon+{\bf W})}$, and use $\exp(\widetilde{\bf W})\mathrm{diag}^{-1}\left(\sqrt{ \sum_f \exp(\widetilde{W}_{f,:})^2 } \right)$ as the model weights.

\section{EXPERIMENTS AND DISCUSSION}
\label{sec:experiments}

\begin{table*}[t]
    \centering
    \caption{Results in terms of validation loss (dev.\ loss) and signal-to-disortion ratio (SDR) in dB on the CHiME2 development and test sets using 100\% (center section) or 10\% (right section) of the training data. $K$ is the total number of layers or iterations, $N$ is the LSTM hidden state dimension or the number of NMF basis vectors, and $P$ is the total number of trainable parameters.}
    \begin{tabular}{ccccc|ccc|ccc}
        & \multicolumn{4}{c}{\bf Architecture} \vline
        & \multicolumn{3}{c}{\bf 100\% of training set} \vline
        & \multicolumn{3}{c}{\bf 10\% of training set}
        \\
        \hline
         & Model & $K$ & $N$ & $P$
         &Dev. loss & Dev. SDR & Eval. SDR
         &Dev. loss & Dev. SDR & Test SDR\\
        \hline
        \multirow{ 6 }{3pt}{\rotatebox{90}{
         Baselines
        }}
        & SNMF, MU & $200$ & 200 & 50k & 0.0987 & 7.14 & 8.18  &
        0.1319 & 6.51 & 7.47  \\
        & SNMF, MU & $200$ & 2000 & 500k & 0.0846 & 7.71 & 8.61  &
        0.0890 & 7.43 & 8.37  \\
        \cline{2-11}
        & LSTM & 2 & 54 & 100k 
        & 0.0408 & 11.51 & 12.53  &
        0.0512 & 10.34 & 11.35  \\
        & LSTM & 2 & 244 & 1M 
        & 0.0339 & 11.95 & 12.90  &
        0.0481 & 10.59 & 11.57  \\
        & LSTM & 5 & 70 & 250k 
        & 0.0426 & 10.90 & 11.94  &
        0.0542 & 10.22 & 11.26  \\
        & LSTM & 5 & 250 & 2.5M 
        & 0.0344 & {\bf 12.35} & {\bf 13.30}  &
        0.0566 & 10.25 & 11.32  \\
        \hline
        \multirow{ 4 }{3pt}{\rotatebox{90}{
         Proposed
        }}
        & DR-NMF & 2 & 200 & 100k 
        & 0.0320 & 10.94 & 11.86  &
        0.0362 & 10.39 & 11.33  \\
        & DR-NMF & 2 & 2000 & 1M 
        & 0.0295 & 11.21 & 12.11  &
        0.0354 & 10.54 & 11.43  \\
        & DR-NMF & 5 & 200 & 250k 
        & 0.0286 & 11.14 & 12.04  &
        0.0332 & 10.88 & 11.80  \\
        & DR-NMF & 5 & 2000 & 2.5M 
        & {\bf 0.0266} & 11.31 & 12.19  &
        {\bf 0.0316} & {\bf 11.12} & {\bf 11.99}  \\
    \end{tabular}
    \label{tab:results}
\vspace{-10pt}
\end{table*}

\begin{figure*}[t]
\centering
\includegraphics[width=0.20\linewidth]{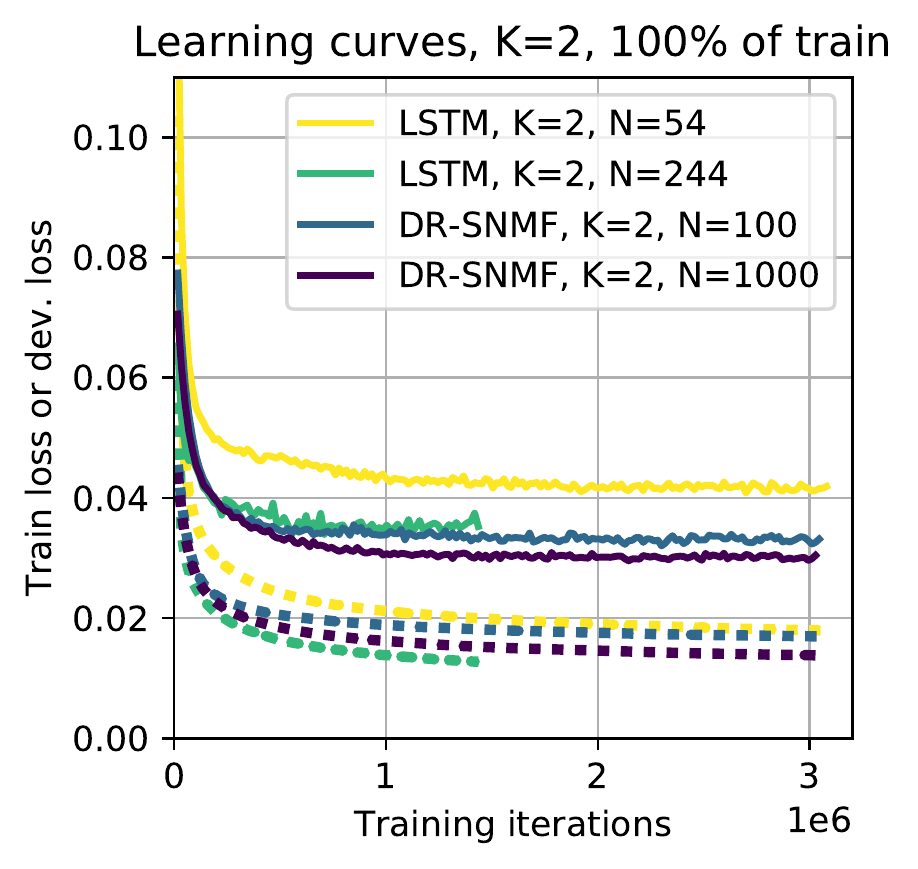}
\includegraphics[width=0.20\linewidth]{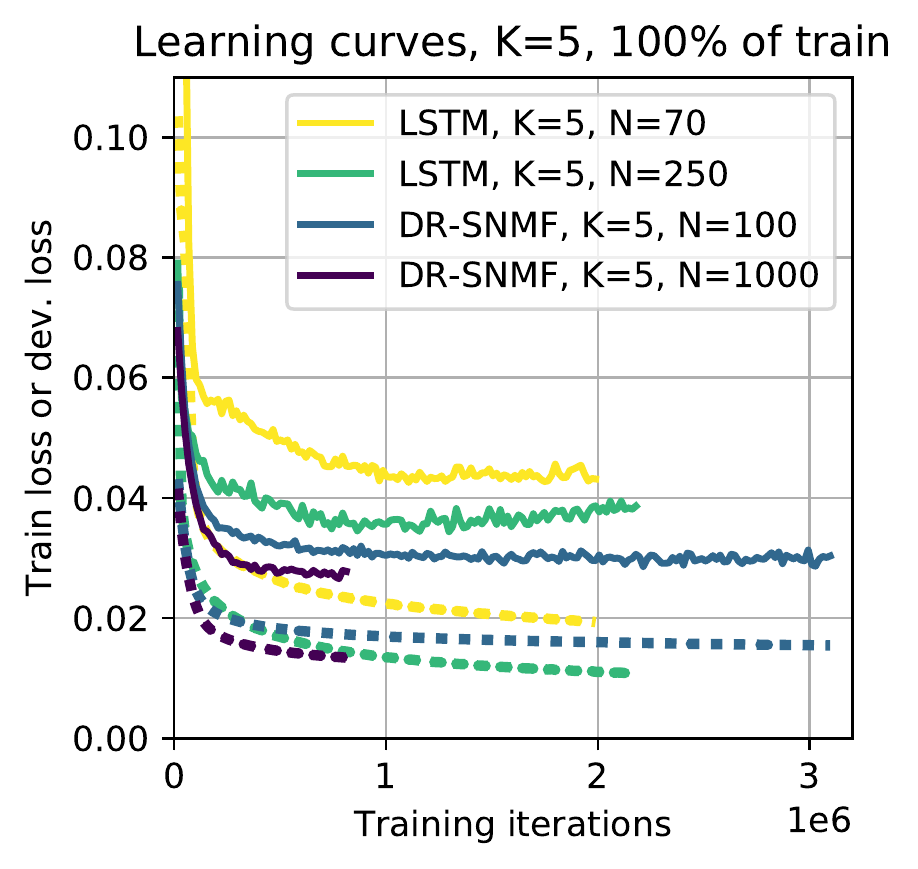}
\vline
\includegraphics[width=0.20\linewidth]{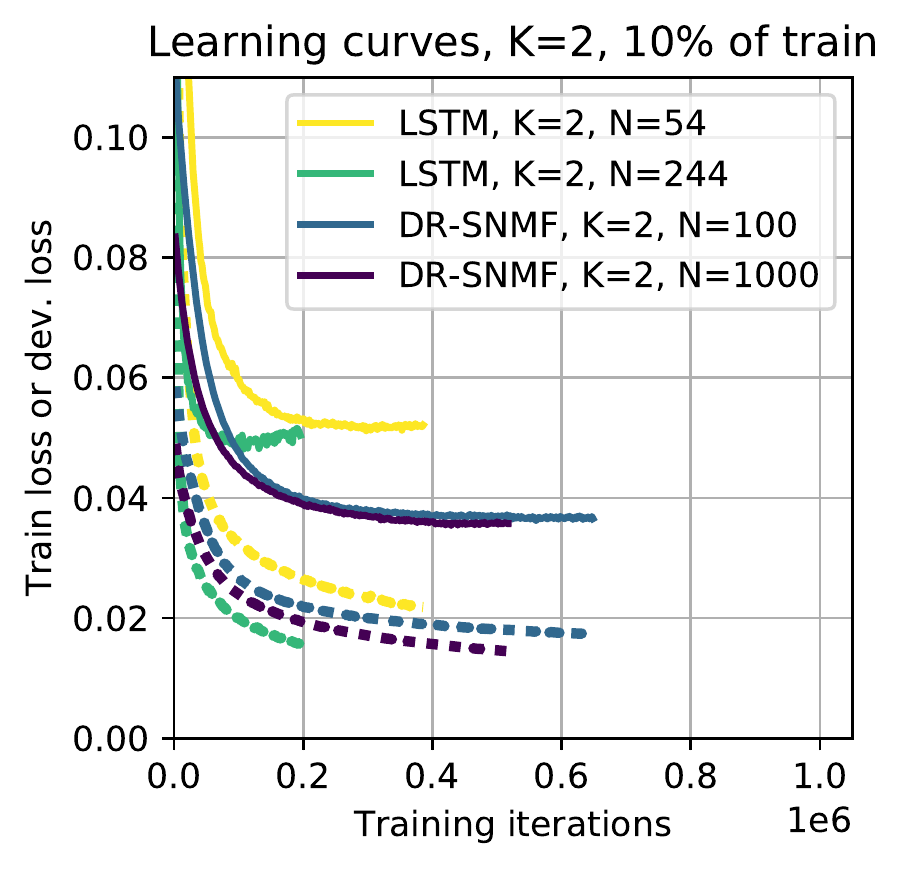}
\includegraphics[width=0.20\linewidth]{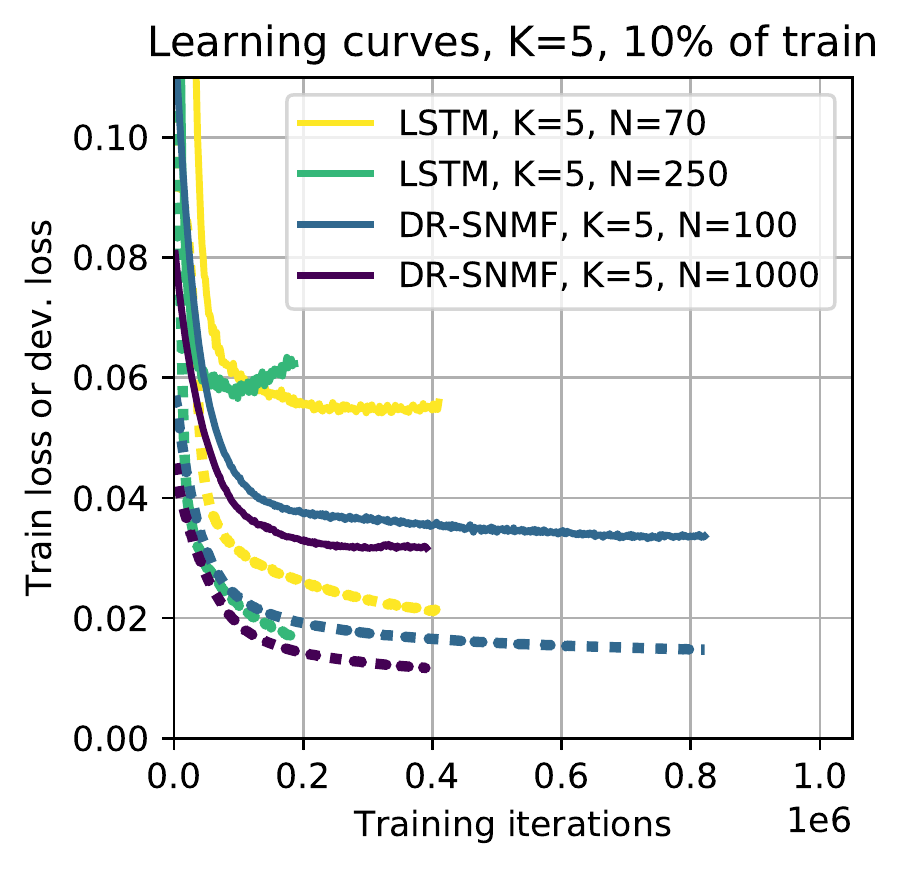}
\vspace{-12pt}
\caption{Learning curves for deep models. Dotted lines are training loss and solid lines are validation loss, where validation loss is computed on the CHiME2 development set. Notice that large LSTM networks generalize well when 100\% of the training data is used (left two panels), but they tend to quickly overfit when only 10\% of the training data is used (right two panels). In contrast, DR-NMF networks achieve good generalization performance and the lowest validation loss using either 100\% or 10\% of the training data (all panels).}
\label{fig:learning_curves}
\vspace{-15pt}
\end{figure*}


For our experiments we use the CHiME2 corpus
\cite{vincent_second_2013} which consists of utterances from the Wall Street Journal (WSJ-0) dataset that are convolved with binaural room impulse responses (RIRs) and mixed with real-world nonstationary noise at six different SNRs from $-6$dB to $9$dB, spaced by $3$dB. The nonstationary noise was recorded in a home environment, and contains a variety of challenging noise types, including music, radio, television, children, and appliances. The RIRs were recorded in the same environment. The training set consists of 7138 utterances, the development set consists of 2460 utterances, and the test set consists of 1980 utterances, which are equally distributed across the six SNRs. All audio is sampled at $16$kHz and we use only the left channel. Separation performance is measured using signal-to-distortion ratio (SDR) from the BSS Eval Matlab toolbox 
 \cite{vincent_performance_2006, vincent_bsseval_website}.

 Two baseline methods are compared to DR-NMF: sparse NMF (SNMF) using well-done multiplicative updates (MU), which is used to initialize the DR-NMF networks, and stacks of LSTM networks. SNMF uses $200$ MU iterations at test time. The depth $K$ of the LSTM stacks are chosen to match the depth of the DR-NMF networks, and the hidden node count $N$ of the LSTMS are chosen to match the counts $P$ of trainable parameters of DR-NMF networks.

 The STFT uses $512$-sample square-root Hann windows for analysis and synthesis with a hop of $128$ samples. Thus, the feature dimension of input spectrogram frames is $F=257$. For training deep models, the input spectrograms are split into sequences no longer than $500$ frames.
 Deep network training uses the Adam optimizer \cite{kingma_adam:_2014} with a batch size of $32$ and default parameters, except for the learning rate. LSTM network training uses gradient clipping to $1$ and a learning rate of $10^{-4}$, and DR-NMF network training uses no gradient clipping and a learning rate $10^{-3}$.
 During training, model weights with the lowest validation loss are saved. Early stopping on the validation loss with a patience of 50 epochs determines training convergence.
 All deep networks are implemented in Keras \cite{chollet_keras_2015} using Theano \cite{theano_development_team_theano:_2016} as a backend. For sparse NMF, we use the Matlab implementation of well-done multiplicative updates
 \cite{le_roux_sparse_2015}. Code to replicate our results is available online\footnote{\url{https://github.com/stwisdom/dr-nmf}}.
 
 The results are shown in table \ref{tab:results}. Notice that when deep models are trained with 100\% of the training set (center section of table \ref{tab:results}), the largest LSTM model achieves the best mean SDR of 12.35 dB on the CHiME2 development set. However, despite not achieving the best SDR, the largest DR-NMF model achieves the best validation loss of $0.0266$. This indicates a discrepancy between SDR, which is computed in the time domain, and the signal approximation loss (\ref{eq:loss}) computed in the magnitude spectrogram domain.
 
 When only 10\% of the training set is used, DR-NMF networks achieve superior performance in terms of SDR and validation loss (right section of table \ref{tab:results}), achieving 11.12 dB SDR on the development set, compared to the best LSTM score of 10.59 dB.
 DR-NMF also outperforms Le Roux et al.'s deep NMF, which achieves 10.20 SDR on the development set using $P=1.2\mathrm{M}$ trainable parameters, $N=2000$ basis vectors, and $K=25$ total MU iterations, the last 3 of which are untied and trained \cite[table 2]{le_roux_deep_2015}.
 
 Figure \ref{fig:learning_curves} shows the learning curves for training the deep networks, which provide insight into the generalization ability of the various models. Notice that large LSTM networks quickly overfit (i.e., the validation loss starts to increase while the training loss continues to decrease) when only provided with 10\% of the training set (right panels of figure \ref{fig:learning_curves}), while DR-NMF networks are consistently robust to overfitting. This suggests that DR-NMF networks exhibit stronger generalization performance compared to LSTMs and thus perform better when provided with less training data. Also, DR-NMF networks achieve the lowest validation loss in all cases.
 
 \vspace{-4pt}
 \section{CONCLUSION}
 \vspace{-2pt}
 
 In this paper, we have proposed a new deep architecture, deep recurrent nonnegative matrix factorization (DR-NMF). This deep network is created by unfolding the iterations of a sequential iterative algorithm, warm start ISTA, that solves the sparse NMF optimization problem. As a result, the DR-NMF network can be initialized with sparse NMF parameters and remains interpretable even after training. Through a speech separation experiment, we showed that DR-NMF networks achieve competitive performance compared to state-of-the-art LSTM networks, consistently exhibiting better generalization in terms of validation loss and yielding better separation performance when a limited amount of training data is available.

\ifjournal
\else
\newpage
\fi



\bibliographystyle{IEEEtran}
\bibliography{waspaa2017}
%
%
%
%
%
%
%
%
%

\end{sloppy}
\end{document}